\documentstyle[prd,aps,tighten,floats]{revtex}

\newcommand{\rf}[1]{(\ref{eq:#1})}

\date{October 5, 1998}

\begin{document}

\author{ M. D. Maia\thanks{maia@fis.unb.br}\\
Universidade de Bras\'{\i}lia, Instituto  de F\'{\i}sica \\
70910-900, Bras\'{\i}lia. D.F.  \\
and\\
Edmundo M. Monte\thanks{edmundo@fis.ufpb.br}\\
Universidade Federal da Para\'{\i}ba\\
Departamento de Matem\'atica\\
58100-000  Campina Grande, Pb. }

\title{The Dynamics of Relativistic Hypersurfaces} 

\maketitle

\begin{abstract}
A dynamical  theory of hypersurface deformations is presented.
It is shown that  a  (n+1)-dimensional space-time can be always foliated by
pure  deformations, governed by a non zero Hamiltonian. 
Quantum deformation states are defined by Schr\"odinger's equation constructed
with the corresponding deformation Hamiltonian operator, interpreted as
the  generator of the deformation  diffeomorphism group.
Applications to quantum gravity and to a modified Kaluza-Klein theory are
proposed.

\end{abstract}


\section{Introduction}
Some present problems in theoretical physics, involve local  changes of 
curvature due to density fluctuations in cosmology,  changes in  the space-time
topology due to black and worm holes formation and involve the
compactification of subspaces in  higher dimensional  models. These problems
motivate the search for an efficient mechanism capable of  describing
classical  and quantum  changes of the geometrical and topological  properties
of   space-time.  
 Considering all possibilities, one such mechanism
would be necessarily too complex. However, for  some applications it is
possible to devise  a simpler mechanism. Take for example
a system of particles distributed in
an neighborhood  of a point of an initial hypersurface  $\bar{V}_{n}$
of  some  $(n+1)$-dimensional manifold  $V_{n+1}$
taken as  a Cauchy surface.  By   heating this system, the particles leave
$\bar{V}_{n}$ but they will not  necessarily  land all into 
another hypersurface later on. This will depend on the energy level, 
interactions and symmetry of the system.   A  simpler  situation
occurs when the particles  have   compatible energies and initial data
so that they move from one surface to another.
In this case, the  evolution of the system, can be
described as a  dynamical deformation of the original hypersurface.

The purpose  of this note is to present a  model of  dynamical deformations of
hypersurfaces and its applications to  quantum gravity in four dimensions and
to a quantum  Kaluza-Klein theory.
The concept of  classical deformation of  a
hypersurface as  a  perturbative process is reviewed  in the next section.
In section 3   some known  geometrical results are applied to show that pure
deformations generate a 
foliation, associated with a dynamical process with a non-zero Hamiltonian.
In section 4  the quantum aspect of  these deformations is discussed and
applied to  quantum gravity described by  3-dimensional hypersurfaces in 
4-dimensional space-time with some limitations.  
In section 5 we extend the  quantum deformations to
higher dimensions obtaining a modified Kaluza-Klein theory.

\section{The Geometry of  Deformations}
Given  a point $p$ in $\bar{V}_{n}$ and  an  arbitrary vector  $\zeta$
in $V_{n+1}$  there is  a  one parameter group of diffeomorphism
 $ h_{s}: V_{n+1}\longrightarrow V_{n+1}$, whose orbit  $\alpha (s)=h_{s}(p)$,
is  an integral curve of  $\zeta$,  passing through  $p$.  
A perturbation of   an object $\bar{\Omega}$ in  $\bar{V}_{n}$
induced  by this
diffeomorphism along $\alpha(s)$  is  given by \cite{Geroch}. 
\[
\Omega =\bar{\Omega}  +s \pounds_{\zeta}\bar{\Omega}.
\]
The hypersurface $\bar{V}_{n}$ containing $p$  is  given by
the local and isometric embedding  $\bar{\cal X}: \bar{V}_{n}\rightarrow
V_{n+1}$ such that\footnote{  
Small case Latin indices $i,j,k...$ refer to an n-dimensional  hypersurface
$V_{n}$ and run from 1 to n. All Greek indices refer to  the
$(n+1)$-dimensional  manifold, running from 1 to $n+1$.
The metric of $V_{n}$ is  denoted by $g_{ij}$  and 
$\nabla_{i}$  denotes  the covariant derivative with respect to
this metric. The covariant derivative with respect to the metric of  $V_{n+1}$
is denoted by  a semicolon and  $\eta^{\mu}_{;i}=\eta^{\mu}_{;\gamma}\bar{\cal X}^{\gamma}_{,i}$
denotes  the projection of the covariant derivative  $\eta^{\mu}_{;\gamma}$
over the submanifold $V_{n}$. }
\begin{equation}
\bar{g}_{ij}=\bar{\cal X}^{\mu}_{,i}\bar{\cal X}^{\nu}_{,j}{\cal G}_{\mu\nu},\;\;
\bar{\cal X}^{\mu}_{,i}\bar\eta^{\nu}{\cal G}_{\mu\nu}=0,\;\;
\bar{\eta}^{\mu}\bar{\eta}^{\nu}{\cal G}_{\mu\nu}= -1,\;\;\bar{k}_{ij}= -{\cal
X}^{\mu}_{,i}\bar{\eta}^{\nu}_{;j}{\cal G}_{\mu\nu}  \label{eq:embeddX}
\end{equation}
where $\bar{\eta}$ is the unit vector normal to
$\bar{V}_{n}$  and $\bar{k}_{ij}$ is  its  extrinsic curvature.
A  deformation of the hypersurface $\bar{V}_{n}$ along $\zeta$  is the subset  of  $V_{n+1}$  described by
the coordinates  ${\cal Z}^{\mu}$ given by
the  perturbation of the embedding vielbein $\bar{\cal X}^{\mu}_{,i}$: 
\begin{eqnarray}
{\cal Z}^{\mu}_{,i}(x^{i},s)&=&\bar{{\cal X}}^{\mu}_{,i}+s\pounds_{\zeta}
\bar{{\cal X}}^{\mu}_{,i} =
\bar{{\cal X}}^{\mu}_{,i}+s[\zeta, \bar{{\cal X}}^{\alpha}_{,i}] \\
\eta & =&\bar{\eta} +s\pounds_{\zeta}\bar{\eta}  =\bar{\eta} +s[\zeta,\bar{\eta}]
\end{eqnarray}
 In order to associate  these deformations with a physical process,
two  conditions are required: In the first place the deformation must
be free of  coordinate gauges. Secondly, the deformed hypersurface must be again
a hypersurface.

Given  two  distinct deformations  of the same $\bar{V}_{n}$
corresponding to two slightly distinct  directions  such that 
$ \zeta'  -\zeta =\delta\zeta $, their  difference is given by 
\[
\delta{\cal Z}^{\mu}_{,i}=
s\pounds_{ \delta\zeta }\bar{{\cal X}}^{\mu}_{,i} =s [\delta\zeta ,\bar{\cal
X}^{\mu}_{,i}].  
\] 
In the theory of elastic membranes the tangent component of the  deformation
tension is canceled by  the assumption that  it is constant, which means
 invariant  under  the diffeomorphisms of the membrane. 
Here  we cannot use the same  argument.
However, if $\bar{V}_{n}$ is  endowed with a general diffeomorphism
group then it is  always possible to find a coordinate system in which
the above Lie bracket vanishes. This means that 
 the deformation can be  canceled or transformed away by a  choice of
coordinate gauge in $\bar{V}_{n}$ which is of course undesirable for
a physically generated deformations. It is possible  to filter out these
coordinate gauges by imposing 
condition on the geometry of $\bar{V}_{n}$ as often done in cosmology
\cite{Bardeen}, 
but such ad hoc  procedure cannot be applied to generic deformations. Again,
comparing with  the   example of elastic membranes, the fundamental modes of
the deformation can  be generated by   
deformations defined along  the direction orthogonal to  $\bar{V}_{n}$.
These  "pure" deformations denoted by  $V_{n}$ are given by \cite{HKT}
\begin{eqnarray}
{\cal Z}^{\mu}_{,i}(x,s) &=&\bar{{\cal X}}^{\mu}_{,i}(x)
+s \pounds_{\eta}\bar{{\cal X}}^{\mu}_{,i} =\bar{{\cal
X}}^{\mu}_{,i}(x)  +s\eta^{\mu}_{,i}(x).
\label{eq:Zi}\\
\eta & = & \bar{\eta} +s\pounds_{\bar{\eta}}\bar{\eta} =\bar{\eta}
\end{eqnarray}  
Once obtained  a coordinate gauge independent deformation we   need to make
sure that it represents a  hypersurface
of  $V_{n+1}$. To see that this is true, the  embedding equations
\begin{equation}
g_{ij}={\cal Z}^{\mu}_{,i}{\cal Z}^{\nu}_{,j}{\cal G}_{\mu\nu},\;\;  
{\cal Z}^{\mu}_{,i}{\eta}^{\nu}{\cal G}_{\mu\nu}=0,\;\;
{\eta}^{\mu}{\eta}^{\nu}{\cal G}_{\mu\nu}=-1  ,\;\;
k_{ij}=-{\cal Z}^\mu {}_{,i}\eta^\nu {}_{;j}{\cal G}_{\mu \nu }
\label{eq:embeddZ}    
\end{equation}
must be   satisfied. Indeed, 
{\em Any  pseudo Riemannian manifold  $V_{n+1} $  with metric  signature 
$(n,1)$ is  necessarily  foliated by pure deformations
of a  given hypersurface  $\bar{V}_{n}$ with  Euclidean signature.} 
 This  follows from  a simple  adaptation of  some well known embedding  theorems
based on manifold deformations \cite{Nash}.
From   \rf{Zi},
we obtain after applying  \rf{embeddZ} the  metric $g_{ij}$ of  $V_{n}$ in
terms of the extrinsic curvature of $\bar{V}_{n}$:  
\begin{equation}
g_{ij}={\cal Z}^{\mu}_{,i}{\cal Z}^{\nu}_{,j}{\cal G}_{\mu\nu}=\bar{g}_{ij}
-2s\bar{k}_{ij} +s^{2}\bar{g}^{mn}\bar{k}_{im}\bar{k}_{jn}. \label{eq:gij}
\end{equation}
On the other hand, from the definition  
$k_{ij}=-{\cal Z}^{\nu}_{,i}\eta^{\mu}_{;i}{\cal G}_{\mu\nu}$, it follows that
\begin{equation}
k_{ij}= \bar{k}_{ij} -s\bar{g}^{mn}\bar{k}_{im}\bar{k}_{jn} \label{eq:kij}
\end{equation}
Comparing with  \rf{gij} we obtain  the  evolution  of the
metric  as\footnote{This is known in the modern literature  related  to 
quantum gravity as  York's expression where  the dot means the derivative with
respect to time.  To maintain the  analogy,  we use  here the same dot
notation:  $\dot{\Omega} =d\Omega/ds$, keeping in mind that  $s$ is  an
independent parameter}  
\begin{equation}
\frac{d g_{ij}}{d s}=\dot{g}_{ij}=-2 {k}_{ij}. \label{eq:YORK}
\end{equation}
Although  \rf{gij} is  exact for  an arbitrary dimension $n$,
its inverse   cannot be calculated exactly.
However, using the matrix notations $\bar{\bf g}=(\bar{g}_{mn})$, ${\bf
g}=(g_{mn})$  and ${\bf k}=(k_{mn})$, the  inverse  metric can be 
given  to any order  $(k)$ of approximation  as
\begin{equation}
\stackrel{(k);\;}{{\bf g}^{-1}}=\left( \sum_{n=0}^{k} (\bar{\bf g}^{-1}{\bf k})^{n}
\right)^{2}\bar{\bf g}^{-1},\;\;\; \; {\bf g} \stackrel{(k)\;\;}{{\bf
g}^{-1}}\approx 1 +0(s^{k+1}) . \label{eq:gk}
\end{equation}
From this we obtain   $\dot{g}_{im}  \stackrel{(k)}{g^{mj}} +g_{im}
\stackrel{(k)}{\dot{g}^{mj}}=0$.  
Therefore,  to any order of approximation  we may
define   $ \stackrel{(k)}{k^{ij}} =\frac{1}{2}
\stackrel{(k)}{\dot{g}^{ij}}$ so that 
\[
\stackrel{(k)\;\;}{k^{mn}}=
\stackrel{(k)}{g^{im}}\stackrel{(k)\;\;}{g^{jn}}k_{mn} 
\] 
In   the  limit $k\rightarrow\infty$ we obtain the exact expression
\begin{equation}
k^{ij}=2\dot{g}^{ij}.
\end{equation}
Thus, in spite of the  approximation  rf{gk}, the   
 indices  of   $k$  are  lowered and risen by the metric of  $V_{n}$,  $g_{ij}$
and $g^{ij}$ respectively  according to \rf{YORK}  
\[
g_{im}g_{jn}k^{mn}\!=\!g_{im}g_{jn}\frac{1}{2}\dot{g}^{mn}\!=\!
\frac{1}{2}\frac{d}{ds}(g_{im}g_{jn}g^{mn})-
\frac{1}{2}\dot{g}_{im}g_{jn}g^{mn}-
\frac{1}{2}{g}_{im}\dot{g}_{jn}g^{mn}= k_{ij}.
\]   
To conclude, we make use of the  particular reference frame adapted to the
foliation, $\eta^{\alpha}=\delta^{\alpha}_{n+1}$ 
where ${\cal G}_{ij}=g_{ij}$, ${\cal G}_{i\,n+1}=0$ and  ${\cal
G}_{n+1\,n+1}=-1$, 
and  the components of the   Riemann tensor 
 are,  after making use of \rf{YORK}
 \begin{equation}
{\cal R}_{\alpha\beta\gamma\delta}{\cal Z}^{\alpha}_{,i}{\cal
Z}^{\beta}_{,j}{\cal Z}^{\gamma}_{,k}{\cal Z}^{\delta}_{,l} 
=R_{ijkl}-  \frac{1}{4}(\dot{g}_{il} \dot{g}_{kj}-\dot{g}_{ik}\dot{g}_{jl}) =
R_{ijkl} - 2k_{i[l}k_{j]k} \label{eq:G} 
\end{equation}
and
\begin{equation}
{\cal R}_{\alpha\beta\gamma\delta}{\cal
Z}^{\alpha}_{,i}\eta^{\beta}{\cal Z}^{\gamma}_{,j}{\cal Z}^{\delta}_{,k}\!\!
=\!\! \frac{1}{2}(\frac{\partial \dot{g}_{ik} }{\partial x^{j}}
-\frac{\partial \dot{g}_{ij}  }{\partial x^{k}}) -\frac{1}{2} 
g^{mn}(\Gamma_{ikm}\dot{g}_{jn}-\Gamma_{ijm}\dot{g}_{kn})
= 2\nabla_{[k}k_{j]i}  \label{eq:C}
\end{equation}
Notice that these tensor equations hold in any arbitrary coordinates and they readily seen
to be  the
integrability conditions for a hypersurface of $V_{n+1}$  \cite{Eisenhart}.
Finally, the  fundamental theorem of
hypersurfaces  states that for  a given pair of tensors  $g_{ij}$ and
$k_{ij}$ satisfying \rf{G} and  \rf{C},  
 there exists a hypersurface  $V_{n}$ embedded  in  ${V}_{n+1}$
described by  ${\cal Z}^{\mu}$ and with normal vector  ${\eta}^{\mu}$
satisfying \rf{embeddZ}. In other words, the  pure deformations given by
\rf{Zi} produce the necessary and sufficient  conditions to generate a hypersurface
through  \rf{YORK}.

\section{Deformation Dynamics}
The next  question concerns the  dynamical  aspect of the  deformations.
To start, we notice that the  first equation  \rf{embeddZ} can be solved as a
tensor equation to give 
$g^{ij}{\cal Z}^{\alpha}_{,i}{\cal Z}^{\beta}_{,j}={{\cal G}}^{\alpha\beta} 
+\Psi^{\alpha\beta}$, where  $\Psi^{\alpha\beta}$ is a non null tensor of
$V_{n+1}$ satisfying the condition ${\cal G}_{\alpha \beta}\Psi^{\alpha\beta}
=-1$. Applying in the  second equation \rf{embeddZ}, it follows that the only
 solution  compatible with ${\cal G}_{\alpha\beta}g^{ij}{\cal
Z}^{\alpha}_{;i}{\cal Z}^{\beta}_{;j}=n$ and ${\cal G}^{\alpha\beta}{\cal
G}_{\alpha\beta}=n+1 $ is 
$\Psi^{\alpha\beta}=-\eta^{\alpha}\eta^{\beta}$, so that 
\begin{equation}
g^{ij}{\cal Z}^{\mu}_{;i}{\cal Z}^{\nu}_{;j}  ={\cal G}^{\mu\nu} -\eta^{\mu}\eta^{\nu}
\label{eq:LEMMA1}
\end{equation}
Using \rf{LEMMA1}, the contractions of  \rf{G}  become
\begin{eqnarray} 
R_{jk} & = & g^{il}R_{ijkl}\!\!=\!\! -(k^{lk}k_{lj}-h k_{jk})+
{\cal R}_{\beta\gamma} {{\cal Z}}^{\beta}_{,j} {{\cal Z}}^{\gamma}_{,k}
+ {\cal R}_{\alpha\beta\gamma\delta} 
{\eta}^{\alpha} {\eta}^{\delta}  {\cal Z}^{\beta}_{,j} {\cal
Z}^{\gamma}_{,k}  \label{eq:Rij}, \vspace{3mm}\\
R & = & -(\kappa^{2} -h^{2})+{\cal R} -2{\cal
R}_{\alpha\beta}\eta^{\alpha}\eta^{\beta}  \label{eq:R}    
\end{eqnarray}
where we have denoted  the   mean curvature of the
deformation  $V_{n}$  by  $h=g^{ij}k_{ij}$ and  $k^{2}=k^{ij}k_{ij}$.  
On the  other hand the contraction of  \rf{C}  gives
\[
\nabla_{k}k^{k}_{i}\! -h_{,i} = {\cal R}_{\alpha\beta}
{\cal Z}^{\alpha}_{,i}{\eta}^{\beta}
\]
Therefore, the   Einstein-Hilbert Lagrangian 
as derived  directly  from   \rf{R}, in arbitrary coordinates is
\begin{equation} 
{\cal L}={\cal R}\sqrt{-{\cal G}}=\left[ {R}+ (\kappa^{2}-h^{2})+2{\cal
R}_{\alpha\beta}\eta^{\alpha}\eta^{\beta} 
\right]  
\sqrt{-{\cal G}}\label{eq:L} 
\end{equation} 
Before  discussing  the Euler-Lagrange equations, let us proceed with  the
definition of the canonical momenta conjugated to the
metric  ${\cal G}_{\alpha\beta}$  with respect to the  deformation parameter
$s$:
\[
\displaystyle
\pi^{\alpha\beta}=\frac{\partial{\cal L}}{\partial
(\dot{\cal G}_{\alpha\beta}) }.
\]
The corresponding  Hamiltonian is given  by the Legendre
transformation 
\begin{equation}
{\cal H}={\cal L}- \pi^{\alpha\beta}\dot{\cal G}_{\alpha\beta}
\label{eq:H}
\end{equation}
where we  notice that   no specific  metric decomposition
(as in the ADM formulation) was used. In generic coordinates  the  
metric  ${\cal G}_{\alpha\beta}$ of  $V_{n+1}$ can be split as 
\[
{\cal G}_{\alpha\beta}=
\left(
\begin{array}{ll}
{\cal G}_{ij} &  {\cal G}_{i\,n+1}\\
{\cal G}_{n+1\,i} & {\cal G}_{n+1\,n+1}
\end{array}
\right),    
\]
Here  ${\cal G}_{ij}$ corresponds to the metric of $V_{n}$ and the  index $n+1$
does not refer to the  deformation parameter  $s$ but rather to  just one of
the   coordinates $x^{n+1}$ of  $V_{n+1}$.
Assuming that the metric $g_{ij}$ of  $V_{n}$ is  written in arbitrary
coordinates,  we may identify without loss of  generality 
${\cal G}_{ij}=g_{ij}$, while keeping the
remaining components  ${\cal G}_{i\,n+1}$ and ${\cal G}_{n+1\,n+1}$ 
completely arbitrary and unspecified.
 Using this notation, the momentum components corresponding to  $g_{ij}$ can be
written as     
\begin{equation}
\pi^{ij} =-( k^{ij}-hg^{ij} +\frac{\partial {\cal
R}_{\alpha\beta}\eta^{\alpha}\eta^{\beta} }{\partial
{\dot g}_{ij}} )\sqrt{-{\cal G}}\vspace{3mm} \label{eq:piij}
\end{equation}
Again, in the particular coordinates adapted to the deformation where
$\eta^{\alpha}=\delta^{\alpha}_{n+1}$,  the scalar  ${\cal
R}_{\alpha\beta}\eta^{\alpha}\eta^{\beta}$   is given by
is  
\begin{equation}
{\cal R}_{\alpha\beta}{\eta}^{\alpha}{\eta}^{\beta} = 
\Gamma^{\alpha}_{n+1\alpha,n+1}-\Gamma^{\alpha}_{n+1\,n+1,\alpha}+
\Gamma^{\beta}_{n+1\,\alpha}\Gamma^{\alpha}_{\beta\, n+1} 
-\Gamma^{\alpha}_{n+1\,n+1}\Gamma^{\beta}_{\alpha\beta} 
= \kappa^{2}-\dot{h} \label{eq:HH}
\end{equation}
Using \rf{YORK}, it follows that 
\[
\theta^{ij} \stackrel{def}{=}\frac{\partial {\cal
R}_{\alpha\beta}\eta^{\alpha}\eta^{\beta}}{\partial k_{ij}}= 2k^{ij}
-\dot{g}^{ij}=0, 
\] 
Now  $g_{ij}$ and  $k_{ij}$ are independent quantities and
$\theta^{ij}$ is  a tensor  so that in any other  coordinate system we  also
have 
\[
\theta'^{ij} = \frac{\partial x'^{i}}{\partial x^{m}}\frac{\partial
x'^{j}}{\partial x^{n}} \theta^{mn} =k'^{ij} -\dot{g'}^{ij}=0
\]
Therefore,  \rf{piij} assumes a more familiar  form in generic coordinates
\begin{equation}
\pi^{ij}= -\frac{1}{2}\frac{\partial {\cal L}}{\partial k_{ij}} =
-( k^{ij} -hg^{ij} ) \sqrt{-{\cal G}}.
\label{eq:pi}
\end{equation}   

Contrarily to  York's relation \rf{YORK}, the pure deformation   \rf{Zi} does not
prescribe   the evolution of   ${\cal G}_{i\,n+1}$ and ${\cal G}_{n+1\,n+1}$.
This means that  the values of the corresponding momenta 
 $\pi^{i\,n+1}$ and  $\pi^{n+1\,n+1}$  remain arbitrary  and
therefore their values should be given  as constraints to the  deformations.
Since  these expressions are all derived from the same 
basic equations \rf{G}, \rf{C} and  \rf{YORK},  they hold along the
entire foliation and as suggested by \rf{HH} we set
\begin{eqnarray}
\pi^{in+1} & = &  -2\frac{\partial{\cal
R}_{\alpha\beta}\eta^{\alpha}\eta^{\beta} }{\partial \dot{{\cal
G}}_{i\,n+1}}\sqrt{-{\cal G}}=0  
\label{eq:DEFCON1},\vspace{3mm}\\ 
\pi^{n+1\,n+1} & = & -2\frac{\partial {\cal R}_{\alpha\beta}\eta^{\alpha}\eta^{\beta} }
{\partial {\dot{\cal G}}_{n+1\,n+1}}\sqrt{-{\cal G}}=0 \label{eq:DEFCON2}
\end{eqnarray}
Equation \rf{DEFCON1}   ensures that the  deformation  does not
 have tangent  components,  a  condition  already  imposed  to guarantee 
a  coordinate gauge independent   deformation. Equation \rf{DEFCON2}
 tells  that  evolution of the system is given by the
parameter $s$ with fixed lapse. 

As  a  consequence of the  above constraints, the indices of
$\pi^{ij}$  may be risen and  lowered with the metric of  $V_{n}$ alone,
without the need to use a supermetric as  required in other formulations \cite{ADM}:
\[
\pi_{ij}={\cal G}_{i\alpha}{\cal G}_{j\beta}\pi^{\alpha\beta} =g_{im}g_{jn}\pi^{mn}= (hg_{ij}-k_{ij})\sqrt{ -{\cal G} }.
\]
Using the definition 
$\pi={\cal G}_{\alpha\beta}\pi^{\alpha\beta}$  and  the  above
constraints  we obtain the familiar expressions
\[
k_{ij} =\frac{1}{\sqrt{-\cal G}} (\frac{\pi}{2}g_{ij}-\pi_{ij}),\;\;\;
 h=\frac{\pi}{2\sqrt{-\cal G}}\;\;\mbox{and}\;\;  k^{2}-h^{2}= \frac{1}{\cal G}(\frac{\pi^{2}}{2}-\pi^{ij}\pi_{ij}).
\]
and  the  Hamiltonian  \rf{H} can be written as
\begin{equation}
{\cal H}= R\sqrt{-{g}} - \frac{1}{\sqrt{-g}}
(\frac{\pi^{2}}{2}- \pi^{ij}\pi_{ij})  -2 {\cal
R}_{\alpha\beta}\eta^{\alpha}\eta^{\beta}  \sqrt{-{\cal G}}. \label{eq:HHH}
\end{equation}
Therefore   Hamilton's  equations describing the foliation of
$V_{n+1}$ with respect to $s$ are
\begin{eqnarray}
\frac{dg^{ij}}{ds} &=&
\frac{2}{\sqrt{{\cal G}}}\left (\frac{\pi}{2}g^{ij}- \pi^{ij} \right)
\sqrt{-{\cal G}},
\label{eq:DOTG}
\vspace{4mm}\\
\frac{d \pi^{ij}}{ds} &=  &
- (R^{ij}-\frac{1}{2}Rg^{ij})\sqrt{g} + \frac{1}{\sqrt{g}}\left[\pi \pi^{ij}-
2\pi^{im}\pi^{j}_{m} 
+\frac{1}{2}(\frac{\pi^{2}}{2}-\pi^{mn}\pi_{mn} ) g^{ij} \right ]
\label{eq:DOTP} 
\end{eqnarray} 
where the  total derivative terms in the right hand side were dropped. 
Notice that the  first  equation coincide  with  York's relation \rf{YORK},
meaning  that the  perturbative process  which  define the  foliation is 
consistent with a  canonical formulation of the theory.

The Euler-Lagrange equations  derived from  \rf{L}  with respect
to  ${\cal G}_{\alpha\beta}$ and to the deformation parameter $s$  gives the
(vacuum) Einstein's equations ${\cal R}_{\alpha\beta}=0$ for  $V_{n+1}$.
If we wish, general relativity  may be implemented by  identifying $s$ as a 
time coordinate and  by imposing the remaining  postulates of that theory.
The identification of  $s$ with   a  coordinate can be made  by  the choosing
 a  foliation based  coordinate system where
$\eta^{\alpha}=\delta^{\alpha}_{n+1}$ so that  we have  a  $(n+1)$-dimensional
general relativity written in a special frame. Of  course, the imposition of
general covariance  eventually will  mix  $s$  with  the remaining coordinates
of $V_{n+1}$ and the undesirable  coordinate gauges  will  appear in the
deformations. 
On the other hand, we may decide to  work with the special frame
where $s$ is  identified with a
coordinate time. In this case the  above  constraints  simply
tell that  Einstein's equations in this  frame assume a  simpler form 
$\partial{\cal L} /\partial g_{i\,n+1}=0$ and $\partial{\cal L} /\partial
g_{n+1\,n+1}=0$, which corresponds to  Dirac's  canonical formulation in a
fixed frame.

\section{Quantum  Deformations}

To the deformation Hamiltonian  ${\cal H}$ given by
\rf{HHH} we may now associate an Hermitian operator  $\hat{\cal H}$ acting on a
Hilbert space of  the wave functions solutions of Schr\"odinger's equation
with respect to  the  deformation parameter  $s$:
\begin{equation}
i\hbar\frac{d\Psi}{ds}= \hat{\cal H}\Psi  \label{eq:SC}
\end{equation}
The solution of this equation   describes what can be called  a quantum 
deformation state  when the usual interpretations of quantum mechanics are
given and when its  semi classical limit  corresponds to a  small 
deformation of the  hypersurface  $\bar{V}_{n}$.  The commutators  involving
$\hat{\cal H}$, $\hat{\pi}^{ij}$ and 
$\hat{g}_{ij}$ correspond to  the relevant Poisson brackets of the
quantum geometry. 

It should be remembered that not all  postulates of general relativity are
being imposed here. However,  it is  useful  to compare the above formulation 
with the  ADM  metric decomposition
procedure  \cite{ADM},
trivially extended to  $n+1$ dimensions. Using the same notation as
before  the metric  of  $V_{n+1}$ decomposes  a  la  ADM  as   
\begin{equation}
{\cal G}^{\alpha\beta}= \left(
\begin{array}{cc}
 g_{ij}-\frac{N^{i}N^{j}}{N^{2}}   &  \frac{N^{i}}{N^{2}}\\                
     \frac{N^{j}}{N^{2}} &  -\frac{1}{N^{2}}   
 \end{array}    \right) \label{eq:ADM}
\end{equation}
where $N$ is the lapse function and  $N_{i}$ the shift  vector  and where $s$
assumes the role of  a coordinate time.
The    Einstein-Hilbert Lagrangian  obtained directly from this metric  is
\begin{equation}
{\cal L}={\cal R}\sqrt{{\cal-G}}= 
  -(k^{ij}-hg^{ij})\sqrt{g}\frac{d g_{ij}}{ds}-(N {\cal H}_{0} 
+ N^{i} {\cal H}_{i})- 2\frac{d}{ds}(h\sqrt{g})-\nabla_{i}\varphi^{i}
\label{eq:H1} .
\end{equation}
Here  $ {\cal H}_{0}=\left[ R-
(k^{2}-h^{2})\right] \sqrt{g} $ is the 
 super Hamiltonian and  ${\cal H}_{i} =\; 2\left[
  \nabla^{j} k_{ji}-h_{,i}\right] \sqrt{g}$
 is  the super momentum.
Excluding the  divergence and total derivatives in \rf{H1}, the effective
Lagrangian  can be written as 
\[
{\cal L}  =\pi^{ij}\frac{dg_{ij}}{ds} -{\cal H}_{ADM}
\]
where $\pi^{ij}= -(k^{ij}-hg^{ij})\sqrt{g}$
is identified with the  momentum  conjugate to  $g_{ij}$   and
${\cal H}_{ADM} = N {\cal H}_{0} +N^{i} {\cal H}_{i}$
is  identified with  the  Hamiltonian. Taking the variation
 of the action    with  respect to  $N$ and $N^{i}$ respectively  we obtain
\begin{equation}
{\cal H}_{0}=(R-(k^{2}-h^{2})\sqrt{g}=0, \;\;\
{\cal H}_{i}=2(\nabla^{j}k_{ij}-h_{,i})\sqrt{g}=0
\end{equation}
so that ${\cal H}_{ADM}=  0 $, up to surface terms.  
It is  a simple matter to see that  ${\cal H}_{0}$ corresponds to the double
trace of \rf{G} and that    ${\cal H}_{i}$ corresponds to the trace of \rf{C}.
Therefore ${\cal H}$ is  constrained to zero over all
hypersurfaces $V_{n}$. This  
is  solved as   Dirac constrained system over hypersurfaces, where$N$ and
$N^{i}$ play the role of the  Lagrange multipliers. 
However,  when  we attempt to canonically quantize the theory, the  constraint
algebra do not propagate  as  expected.
In fact, the  Poisson bracket algebra,   
\[
\begin{array}{ll}
\left[ {\cal H}_{i}(x),{\cal H}_{j}(x')\right]  =  
-{\cal H}_{j}(x)\frac{\partial\delta(x,x')}{\partial x'^{i}} +{\cal H}_{i}(x)
\frac{\partial \delta(x,x')}{\partial x'^{j}}\vspace{3mm}\\
\left[ {\cal H}_{i}(x),{\cal H}_{0}(x') \right] =\;\; {\cal H}_{0}(x)
\frac{\partial\delta(x,x')}{\partial x'^{i}}\vspace{3mm}\\ 
\left[ {\cal H}_{0}(x),{\cal H}_{0}(x') \right]  =  g^{ij}(x){\cal H}_{i}(x)
\frac{\partial \delta(x,x')}{\partial x'^{j}} - g^{ij}(x'){\cal
H}_{i}(x')\frac{\partial \delta(x,x')}{\partial x'^{j}} 
\end{array}
\]
does not remain closed as a  Lie algebra from one hypersurface  to
another \cite{KK:1,Isham,KK:2}.  

The ADM  formalism can also be described in terms of a parametric
foliation of $V_{n+1}$ generated by  deformations of  hypersurfaces  along  the
direction $\zeta$ constructed with the components $N$ and $N^{i}$.  However,
due to the general covariance of $V_{n+1}$,  the  tangent component  of this  deformation
characterized by the shift vector  $N_{i}$ cannot be dispensed with.
As  we have already seen,   this
 implies in the emergence of  coordinate gauges
and  that the  Hamiltonian becomes constrained to zero.  
We notice again that a basic difference from the  present hypersurface dynamics
and the  ADM formulation  is that  the
deformation parameter  $s$ is not a coordinate and therefore it is not
subjected to  general coordinate transformations. As it was shown by Dirac, if
we chose specific coordinates in  the  ADM formulation then  the constraint on
the Hamiltonian is also removed. However, admittedly  it appears to be  very
difficult to conciliate general covariance with  a  non constrained canonical
formalism of general relativity \cite{KK}. 
A possible solution for this conflict  may be obtained using another theory
in which the  postulates concerning symmetries are different from those
required in general relativity. In the next section  we  examine one of these
possibilities.

\section{Quantum Kaluza Theory}

Kaluza-Klein theory represents a very intuitive scheme for the unification  of
fundamental interactions based on the Einstein-Hilbert principle.
Unfortunately, it has  become somewhat stagnant in face  of
two major difficulties. 
One of these  problems   is  the  inability  of the theory to
generate  the light chiral  fermions that are  expected to be present at the
electroweak level.  The second and   perhaps
more serious problem is that it  inherits the non-renormalizability of
Einstein's general relativity  \cite{KK}. 
In that theory, dimensional reduction   uses the  spontaneous compactification of the extra
dimensions so as to render them unobservable at the lower energy levels. 
As   a  result,  the  physical space  would have a topology
like $V_{4}\times B_{N}$  where $B_{N}$  is some  compact space with a maximal
number of  Killing vector fields\footnote{  $M_{4}$ denotes  Minkowski's
space-time. $V_{D}$  is  a 
higher dimensional taken to be the physical space  solution of the vacuum
Einstein's equations, with  $D\ge 4$. Capital Latin indices run from 4 to
$D_4$ }.
Unlike  general relativity, the ground state of the theory  $ M_{4}\times
B_{N}$  is not flat.  A theorem due to   Lichnerowicz implies that in this
case the  masses  of fermions  would  be  proportional to the inverse of the
curvature radii of the compactified dimensions\footnote{The old 5-dimensional
theory  does not present 
the  same   problem because its ground state   $M_{4}\times S^{1}$ happens to
be flat. However as we know today, that theory  does not  have sufficient
degrees of freedom  to promote the intended unification of  gravitation with
electromagnetism.}.  Actually,  we would  end  up   with
an   infinite  tower of   massive states in four dimensional
space-time  which could still be present today \cite{Kolb}. 
Many different proposals have been presented to modify the  non-Abelian
Kaluza-Klein theory, but   without a satisfactory solution
of the above problems \cite{Weinberg,Wett,RS,MaiaKK,Wesson,RT}. 
In this section we  apply the   deformation program to  a non-compact
version of Kaluza-Klein theory where the  compactification hypothesis 
is replaced by the more general
concept of  breaking the translational symmetry along the extra dimensions.

Again we assume that   physical space  $V_{D}$, $D\ge 4$
is  a  D-dimensional pseudo-Riemannian manifold, with  metric signature
$(3,D-3)$, solution of the $D$-dimensional vacuum   Einstein's
equations  ${\cal R}_{\alpha\beta}=0$.  
 Dimensional reduction corresponds to a  loss  of  energy of the system and
consequently a loss of some  of these degrees of freedom. It is reasonable to
suppose that this  happened in sequence,  one after the other. 
That is, $V_{D}$  reduces to  $V_{D-1}$, then to
$V_{D-2}$... and so on, till reaching  the present  day $V_{4}$ which remains as
a submanifold of $V_{D}$.
In each of these steps,  the particles of the reduced system, stay at a
certain hypersurface $V_{n} $  of $V_{n+1}$, for  $n=4,5,\cdots D-1$
so that we may apply the  deformation dynamics of last sections.
To complete,  we follow the same  principle of general  relativity  to stablish
the ground state to be  the D-dimensional   Minkowski space  $M_{D}$, instead of
the traditional $M_{4}\times B_{N}$. 
With this choice  the consequences of the compactification
are removed and in principle  large  mass  fermions will not
appear. 

In the  original Kaluza-Klein theory the total amount of  energy  required to
compactify the  extra degrees of freedom down to Planck's wavelength of
$10^{-33}$ cm   is  Planck's  energy  $\approx 10^{19}$-Gev proportional to
the  ratio of  the coupling constants  in  the  unified Lagrangian.
Therefore, it is reasonable  to suppose here that this  corresponds to the
energy lost in the breaking of the translational  symmetry.
Reciprocally,  the same of energy  is required to restore that symmetry.
Therefore  the range of definition of the  theory  is    that of  Planck's
energy, which definitively  denounces a quantum affair.
In accordance with this, we use  quantum deformations to  access the extra
dimension $\eta^{A}$ by the translation generated  by  the deformation
Hamiltonian, according to  Schr\"odinger's equation 
corresponding to  $s^{A}$: 
\begin{equation}
i\hbar\frac{d\Psi_{A}}{d s^{A}} =\hat{\cal H}_{A}\Psi_{A}, \;  \;\;
A=n+1...D\label{eq:SCA} 
\end{equation}
where the operator $\hat{\cal H}_{A}$ corresponds to the deformation
Hamiltonian  along 
the extra  dimension $\eta_{A}$. The corresponding dimensional reduction
occurs when  $\hat{\cal H}_{A}$ becomes  constrained to zero. In the
application to  general relativity this  results from the  imposition of general covariance
to  the extra dimensions  $\eta_{A}$  but in a Kaluza-Klein scheme this condition
is not required.
This represents a  contrast with  Klein's compactification hypothesis
which enabled the harmonic expansion of the fields in terms of the internal
variables, producing a hybrid theory where a quantum sized compact geometry was
described by a classical metric.  

The  choice of  ground state $M_{D}$  and its  signature $(3, D-3)$ implies
that the  complementary space orthogonal to the space-time is Euclidean,
 with a compact internal group  $SO(D-4)$. 
From the embedding point of view this seems to be too restrictive.
In fact, it is   well known  that  any  four dimensional
space-time can be  analytically embedded in  $M_{10}$. 
However, the  theorem  of last section dispenses  with
the analyticity of the embedding (which in terms of  high energy physics sounds
like a  luxury anyway) and we  have in fact differentiable embeddings.
 This  increases the upper limit of D to  14 dimensions
\cite{Greene}, making the model compatible with an  $SO(10)$ gauge group.
The use of  Euclidean signature in the orthogonal space may also be  a cause
for concern  because of 
 the  dependence of the fermion chirality  on the signature of
the space \cite{Wett}. 
Yet, we   must keep in mind the possibility  of  that the signature of that
space may change as a result of the quantum dynamics.

To end,  we  will  write the equivalent  
 to the classical  Kaluza-Klein  metric. Consider the
 ground state  $M_{D}$ containing the  space-time
$V_{4}$ as  a subspace and that   we have sufficient energy so that all
dimensions are freely accessible. Since  we have no memory  on which
dimension was reduced first and  which one  should be  restored first,
 we  define a single direction $\eta$ orthogonal to $V_{4}$ as a
linear combination  of all  $(D-4)$  extra dimensions:
\[
s\eta =\sum_{5}^{D-4}  s^{A}\eta_{A},\;\;  s\neq 0
\]
The pure deformation along this direction with parameter $s$ is given by 
\[
{\cal Z}^{\mu}_{,i}  =\bar{\cal X}^{\mu}_{,i} +s\eta^{\mu}_{,i}=\bar{\cal
X}^{\mu}_{,i} +\sum_{A}x^{A}\eta^{\mu}_{A,i} .
\]
This  corresponds to the  superposition of
$D-4$  pure deformations, one for each normal $\eta_{A}$ given by \rf{SCA}.
In the case of the chosen signature, the  isometric embedding equations for
each of these deformations are now given by 
\begin{equation}
g_{ij}={\cal Z}^{\mu}_{,i}{\cal Z}^{\nu}_{,j}{\cal G}_{\mu\nu},\;\;
{\cal Z}^{\mu}_{,i}\eta^{\nu}_{A}{\cal G}_{\mu\nu}=s^{M}A_{iMA},\;\;
{\eta}^{\mu}_{A}{\eta}^{\nu}_{B}{\cal
G}_{\mu\nu}=g_{AB}=-\delta_{AB}   \label{eq:multi}
\end{equation}
where
\begin{equation}
k_{ijA}=-{\cal 
Z}^{\mu}_{,i}{\eta}^{\nu}_{A;j}{\cal G}_{\mu\nu},\;\;
A_{iAB}=\eta^{\mu}_{A,i}\eta^{\nu}_{B} {\cal G}_{\mu\nu},   \label{eq:embedZZ}
\end{equation}
As compared with \rf{embeddZ}, we  notice the  emergence of a new  geometrical
object,  the twisting vector $A_{iAB}$, which appears in the derivative  of
$\eta_{A}$ 
\[
 \eta^{\mu}_{A,i}=  -{g}^{mn}{k}_{im}{\cal Z}^{\mu}_{,n}
+g^{MN}{A}_{iMA} 
 \] 
Equations \rf{multi}  can also be seen as  an expression of the  metric
components of  $M_{D}$  written in 
terms of the coordinate  basis  defined  by the multiparameter deformation. In
order to  make a distinction from  ${\cal G}_{\alpha\beta}$ we denote it by
${\cal \gamma}_{\alpha\beta}$ with separate components:
\[
\begin{array}{lll}
{ \cal \gamma}_{ij} &= &{\cal Z}^{\mu}_{i,}{\cal Z}^{\nu}_{,j}{\cal G}_{\mu\nu}=
\bar{g}_{ij} -2s^{A}\bar{k}_{ijA}
+s^{A}s^{B}\left ( \bar{g}^{mn}\bar{k}_{imA}\bar{k}_{jnB}
+\bar{g}^{MN}A_{iMA}A_{jNB}\right )\\ 
{\cal \gamma}_{iA} & =& {\cal Z}^{\mu}_{,i}\eta^{\nu}_{A}{\cal G}_{\mu\nu} =
s^{M}A_{iMA}  \\
{\cal \gamma}_{AB} & =&\eta^{\mu}_{A}\eta^{\nu}_{B}{\cal G}_{AB} =g_{AB}
\end{array}
\]
or,  after   denoting
\[
\begin{array}{ll}
g_{ij} =\bar{g}_{ij} -2s^{A}\bar{k}_{ijA} +s^{A}s^{B}\bar{g}^{MN}
\bar{k}_{imA}\bar{k}_{jnB} \\
  A_{iA}  =s^{M}A_{iMA}
\end{array}
\]
this metric can be written as  
\begin{equation}
{\cal \gamma}_{\alpha\beta} =
\left(
\begin{array}{cc}
           g_{ij}+ g^{MN}A_{iM}A_{jN} & \;\;\;A_{iA}\;\; \\
		          A_{jB}          & \;\;\; g_{AB}\;\;
\end{array} \right) \label{eq:KK}
\end{equation}
which has the same appearance as the Kaluza-Klein metric ansatz, but in fact
it contains some relevant differences:
Firstly, notice that the  metric  $g_{ij}$ is not the metric of the  initial
hypersurface but rather the metric of a deformation. The other relevant
difference is that  although there are no  compact spaces, the twisting vector
 $A_{iAB}$  transforms like  Yang-Mills  potential over  the  space-time,
relative to the  group  of  isometries of the  orthogonal  space
\cite{MaiaEdm}. The metric $g_{AB}=-\delta_{AB}$  is  the  metric  of the
extra dimensional space  which is naturally written in its Killing basis of the
group   $SO(D-4)$. Finally,  the expansion of  the Lagrangian 
calculated from this metric in terms of  $s^{A}$  produces 
\[
{\cal L } ={\cal R}\sqrt{-{\cal \gamma}}  ={R}\sqrt{-{g}}
+\frac{1}{4}tr {F}^{2}\sqrt{-{g}} 
+\;\;\mbox{extra terms} 
\]
where ${F}^{2}={F}_{\mu\nu}{F}^{\mu\nu}$ and ${F}_{\mu\nu}$  is
constructed with the  potentials  ${A}_{iA}=s^{A}{A}_{iMA}$ of
$V_{n}$  \cite{MaiaMeck}.

Comparing   \rf{KK}   with the  ADM  metric we also notice an 
analogy  between  the Yang-Mills potential  $A_{iAB}$ and the shift vector
$N^{i}$. 
However, this is  only apparent because   contrarily to  \rf{ADM}, the
deformation   given by  \rf{KK} is 
pure  so that it is  coordinate gauge free and the dynamical system can be
constructed in a non constrained way. On the other hand  the  deformation
implicit  in \rf{ADM}  contains a genuine  transverse component associated with
the  shift  vector  $N_{i}$. 

\vspace{1cm}

In resume,  we  have started with perturbative  deformations of a hypersurface
which  is  consistently associated with a classical
dynamical process. Since the corresponding canonical  formulation is not
constrained, the system can be quantized in a straightforward manner, where the
fundamental modes are solutions of   Schr\"odinger's  equation relative
to the deformation parameter and to the deformation Hamiltonian.

As a possible application, we considered general relativity as  described by
3-dimensional hypersurface deformations with respect to  an independent
parameter $s$ regarded as a time parameter in the
sense of Liebnitz,  characterized within each  dynamical process. 
In this case, the  quantum deformations could be applied to general
relativity by  using a coordinate transformation 
to the system adapted to the foliation, construct  ${\cal H}$ and then 
transform back to  general coordinates \cite{MaiaTime}.

However, in the four dimensional formulation of  general relativity it appears
to be  no room for an extra time parameter and at the  end, for  practical purposes we need
to identify  $s$  with a coordinate time. The result is similar
to the well known  canonical formulation  of  the gravitational field in a
fixed frame.  Thus, in
the context of  general relativity  this approach  has some objectionable
limitations and it cannot be  generally acknowledged as  a  valid  solution to
the quantum gravity problem \cite{KK}.   

Perhaps a more  realistic  alternative is to consider  the
deformation parameter as a independent variable in a higher dimensional
theory such as a modified version of Kaluza-Klein theory. This   has the
advantage over general relativity of not imposing general covariance on the
extra dimensions while keeping all properties of the four dimensional
space-time intact. 

In this  quantum Kaluza-Klein  model based on  deformation dynamics the extra
dimensions are  physically  accessible by  translational motions at  high
energies, generated by the   deformation Hamiltonian operator in  
Schr\"odinger's equation 
\[
i\hbar\frac{\partial \Psi}{\partial s} =  \hat{\cal H}\Psi
\]
where the  resulting  quantum  state of the deformation is a 
superposition of $D-4$ independent states $\Psi_{A}$. The result is
  quantization of the  four dimensional  geometry
 as opposed to the  usual three dimensional  approach \cite{Coleman,Gibbons},
 with the advantage that no  restrictions are imposed on  general relativity. 
Moreover, the  resulting  multiparameter quantum deformation can produce
signature and topological changes, which  are  
relevant  for  a the  expected results of a quantum theory of gravity
\cite{Hawking,Dowker,Balachandran}.

\end{document}